\documentstyle[11pt,mrs2003,epsfig]{article}

\begin{document} 

\title{POLARIZATION MEASURES AND NATURE OF DARK ENERGY}

\author{L.P.L. Colombo, R. Mainini \& S.A. Bonometto}

\affil{Physics Dep. G. Occhialini, Universit\`a degli Studi di
Milano--Bicocca, Piazza della Scienza 3, I20126 Milano (Italy) \&
I.N.F.N., Via Celoria 16, I20133 Milano (Italy)}

\begin{abstract} 
Polarization measures, on wide angular scales, together 
with anisotropy data, can fix DE parameters. Here we discuss
the sensitivity needed to provide significant limits. 
Our analysis puts in evidence that a class of models predicts 
low correlation or anticorrelation between polarization and anisotropy 
at low $l$. This class includes open models and models with DE due to 
a Ratra--Peebles (RP) potential. Results on this point, given in a 
previous paper of ours, are updated and partially corrected. 
We outline that, with the sensitivity of experiments like SPOrt 
or WMAP, high values of $\Lambda$ (energy scale in the RP potential) 
can be excluded. With the sensitivity expected for PLANCK, the
selection will extend to much lower $\Lambda$'s.

\end{abstract} 
 
\section{Introduction} 
The nature of Dark Energy (DE) is one of the main puzzles of cosmology.
DE was first required by SNIa data [1], but a {\it flat} Universe with 
$\Omega_m \simeq 0.3$ and $\Omega_b h^2 \simeq 0.02$ is also favored 
by CBR and LSS observations [2,3] ($\Omega_{m,b}$: matter, baryon 
density parameters; $h$: Hubble parameter in units of 100 km/s/Mpc; 
CBR: cosmic background radiation; LSS: large scale structure). 

DE could be a false vacuum; then, its pressure and energy density 
($p_{DE}$ and $\rho_{DE}$) have ratio $w=-1$. This however requires
a severe fine tuning at the end of the EW transition.
Otherwise, DE can be a scalar field $\phi$ 
self--interacting through a potential $V(\phi)$
(dynamical DE [4,5,6]). Then
$$
\rho_{DE} = {{\dot \phi}^2/ 2 a^2} + V(\phi)~~,~~~~
p_{DE} = {{\dot \phi}^2/ 2 a^2} - V(\phi)
\eqno (1.1)
$$
(derivatives are in respect to the conformal time $t$). 
As soon as $\rho_k = {\dot \phi}^2/2 a^2 < V$, it is $w<0$
For $\rho_k/V \simeq 1/2$, it is $w \simeq -1/3$ and dynamical DE 
approaches an open CDM behavior. Smaller $\rho_k/V$ ratios 
approach $w= -1$ and a $\Lambda$CDM behavior. To work 
out $w(a)$, the Friedman equations, together with the equation of 
$\phi$, are to be integrated; the solutions depend on the shape of 
$V$, which, in principle, is largely arbitrary. 

Among potentials admitting a {\it tracker} solution, the RP [5] and 
SUGRA [6] expressions
$$
V(\phi) = \Lambda^{4+\alpha}/\phi^\alpha
~,~~~~~~~~~
V(\phi) = (\Lambda^{4+\alpha}/\phi^\alpha) \exp (4\pi \phi^2/m_p^2)~,
\eqno (1.2)
$$
are particularly relevant, as they
originate within the frame of Supersymmetric (SUSY) theories. 
Here, $\Lambda$ is an energy scale, set in the interval $1$--$10^{12}\, 
$GeV; $m_p$ is the Planck mass. Once $\Lambda$ and $\Omega_{DE}$ are 
fixed, the exponent $\alpha$ is set. RP and SUGRA potentials yield 
fast and slowly varying $w$, respectively.

Dynamical DE and $\Lambda$CDM often predict similar observational 
outputs. This is welcome, as $\Lambda$CDM is a good fit to data. 
However, here we show that measures of 
anisotropy and polarization of CBR, at large angular scales,
constrain $V$ and distinguish dynamical DE from
$\Lambda$CDM. Accordingly, experiments in progress
already exclude some parameter range for RP models and
higher sensitivities can discriminate even better.
In a previous paper [7], similar 
results were provided; that paper, however, contained a numerical 
mistake. Here we correct some of its quantitative results.

Our procedure includes a likelihood analysis, assuming polarization
data provided by the Sky Polarization Observatory [8], 
both with the expected experimental noise level and a 
higher sensitivity. Results, however, can be straightforwardly extended 
to other observational contexts.

\section{CMB angular spectra from the Boltzmann equations; theory}
The angular CBR spectra $C_l^{T,E,TE}$ (only $E$--mode
is considered through this paper) can be worked out from
the linear fluctuation evolution, obtained by solving
the Boltzmann equation for the photon distribution. The treatment
is discussed in a number of papers [9], giving also the equations 
for the other components of a model. In this section we show that
these equations, in several cases, yield a low or negative
$C_l^{TE}$ for low $l$. All definitions used are the 
same as in CMBFAST [10].

This effect arises because of the simultaneous action of the ISW effect and 
of the opacity $\tau = \int_t^{t_o} n_e(t') \sigma_T a(t')\, dt'$. (Notice
that $-{\dot \tau}=a\, n_e \sigma_T $); $n_e$ and $\sigma_T$ are the 
free electron density and the Thomson cross--section. ISW effect arises 
when we pass either from matter to curvature dominance (open models) 
or from matter to vacuum dominance ($\Lambda$CDM models). 
However, only in the former case and in the presence of opacity, 
anticorrelation arises. Some RP models induce anticorrelation because 
their features more closely approach open CDM, rather than $\Lambda$CDM.

Let us then indicate by $F_l(k,t)$ and $G_l(k,t)$ the 
Boltzmann components for anisotropy and polarization
($k$ is the wave--number). 
The equations for the $G_l$ components in flat models read:
$$
\dot G_l = -{\dot \tau} \big[-G_l + {\Pi \over 2}\big( \delta_{l0}
+{\delta_{l2} \over 5}\big)\big] + {k \over 2l+1} [l\, G_{l-1}
- (l+1) G_{l+1} ] ~
\eqno (2.1f)
$$
($\delta_{ln}$ is the Kronecker symbol). Here
$$
\Pi = G_o + G_2 + F_2,
\eqno (2.2)
$$
is the only vehicle from anisotropy to polarization. 
In open models, eq.~(2.1f) becomes:
$$
\dot G_l = -{\dot \tau} \big[-G_l + {\Pi \over 2}\big( \delta_{l0}
+{\delta_{l2} \over 5}\big)\big] + {\beta \over 2l+1} [l b_l\, G_{l-1}
- (l+1) b_{l+1} G_{l+1} ] ~;
\eqno (2.1o)
$$
here $\beta^2 = k^2 + K$ and $b_l^2 = 1 - K l^2/\beta^2$, with $K =
-(1-\Omega_m) H_0^2$, $H_0$ is today's Hubble parameter.

Initially all $G_l$ terms are zero; to switch them on, the quadrupole 
$F_2(k,t)$ must be great when $n_e$ is not so low.
For wavelengths $2\pi/k$ entering the horizon well after recombination,
$F_2(k,t)$ switch on when $n_e$ has almost vanished,
unless reionization occurs. Notice that the horizon size at
recombination corresponds to $l \ll 200$. Without reionization, below 
such $l$, we expect low $G_l$'s.

Let $F_l^o$ ($G_l^o$) be the present value of harmonics. In spatially 
flat models, the angular spectra read
$$
C^T_l = {\pi \over 4} \int d^3k\, P_o(k) \, \left|{F_l^o(k) }\right|^2 ~, ~
C^P_l = {\pi \over 4} \int d^3k\, P_o(k) \, |G_l^o(k)|^2 ~,~
C^{TE}_l = {\pi \over 4} \int d^3k\, P_o(k) \, F_l^o(k) G_l^o(k) ,
\eqno (2.3f)
$$
$P_o(k)$ being the primeval fluctuation spectrum. In open model,
instead, they read
$$
C^T_l = {\pi \over 4} \int d^3\beta\, P_o(q) \, \left|{F_l^o(\beta)}
\right|^2 ~, ~
C^P_l = {\pi \over 4} \int d^3\beta\, P_o(q) \, |G_l^o(\beta)|^2 ~,~
C^{TE}_l = {\pi \over 4} \int d^3\beta\, P_o(q) \, {F_l^o(\beta)} G_l^o(\beta) ~;
\eqno (2.3o)
$$
here $q= (\beta^2-4K)^2/\beta(\beta^2 -K)$. Clearly, for $\Omega_m=1$,
both $\beta$ and $q$ return $k$.

Comparing eqs. (2.1) and (2.3) shows suitable shifts
in the $k$--space. In particular, the $b_l$ coefficients cause
a shift of $C_l$ peaks, while the passage from $k$ to
$\beta$ and $q$, in eq.~(2.3o), displace the power through 
the harmonics $F_l$ at small $l$, i.e. on scales 
comparable with the curvature scale. 
Apart of these shifts, the gravitational field fluctuations 
($\dot h$ and $\eta$) obey similar but different equations:
$$ {\rm Open~models:} ~~~~~~~~~~~~~~~~~~~~~
2{\bar k}^2 {\dot \eta} = 8\pi G a^2 [\sum_c (\rho_c+p_c) \theta_c 
- {\dot h} \rho_{o,cr} (1-\Omega_m)/a^2] .
\eqno (2.4o)
$$
$$ {\rm Flat~models:} ~~~~~~~~~~~~~~~~~~~~~~~
2k^2 {\dot \eta} = 8\pi G a^2 [\sum_c (\rho_c+p_c) \theta_c + 
(\rho_{DE}+p_{DE}) \theta_{DE}] .
\eqno (2.4f)
$$
Here $\sum_c$ sums over all (relativistic or non--relativistic) 
matter components apart of DE; $\rho_{o,cr}$ is today's critical density.
The wave--numbers $\bar k^2$ and $k^2$ are just shifted by 
$3 K$.

Notice that, in a $\Lambda$CDM model, no DE fluctuations 
exist while $\rho_{DE} = - p_{DE}$, so that the second 
term at the r.h.s. of eq.~(2.4f) vanishes.
In models with dynamical DE, instead, $\theta_{DE} \neq 0$
and $\rho_{DE} \neq - p_{DE}$. 
Accordingly, the second term in square brackets in eq.~(2.4f) may 
read $\theta_{DE} \rho_{o,cr} (1-\Omega_m)(1+w) (\rho_{DE}/\rho_{o,DE})$ 
(the last parenthesis tells us how DE energy scales with $a$). 
This term is analogous to the second term in square bracket in eq.~(2.4o)
and would coincide with it if $w=-1/3$ and, namely, if $\theta = -{\dot h}/2$.
If this is true, apart of a different power 
distribution along the $l$ axis and geometric effects at greater $l$,
there can be similarities in the behavior of open and dynamical 
DE models.
The relation between $\dot h/2$ and $\theta_{DE}$ can then be studied
through the equation
$$
\theta_{DE} + {{\dot h} \over 2} = - {1 \over 1+w}\, \big[{\dot \delta_{DE}}
+ 3 {\dot a \over a} (c_s^2-w) \delta_{DE}\big],
\eqno (2.5)
$$
whose validity indicates that DE behaves as a fluid.
An order of magnitude estimate, however, tells us soon that $\theta_{DE} 
\sim k^2 t\, \delta_{DE}$; then, the ratio between $\theta_{DE}$ and 
the r.h.s. is $\sim (t/L)^2$, where $L$ is the scale related to $k$. 
Before horizon crossing ($t \ll L$), the $\theta_{DE}$ term is 
negligible, in comparison with the r.h.s.. Hence, $\dot h/2$ 
equates the r.h.s. and, therefore, the ratio $-\dot h/2 
\theta_{DE}$ exceeds unity. At horizon crossing such ratio must 
approach unity and keep such as $t$ grows greater than $L$.
The main differences between open and dynamical DE models, in the r.h.s. 
of eqs.~(2.4) are therefore relegated to times before horizon crossing. 
Afterward, the residual difference is due to a factor $1+w$. 

The changes in $\dot \eta$ directly act on $F_2$. The equation fulfilled 
by this spectral component reads:
$$
\dot F_2(k,t) = -\dot \tau \left[-F_2(k,t) 
+ {\Pi(k,t) \over 10}\right]  
+ {k \over 5} [2\, F_1(k,t) - 3 F_3(k,t) ]
+{8 \over 5} \left({\dot \eta(k,t)}
+ {{\dot h(k,t)} \over 6} \right)
\eqno (2.6f)
$$
in flat models, while in open models we have:
$$
\dot F_2(\beta,t) =  -\dot \tau \left[-F_2(\beta,t) 
+ {\Pi(\beta,t) \over 10}\right] 
+ {\beta \over 5} [2 b_2\, F_1(\beta,t) - 3 b_3 F_3(\beta,t) ]
+{8 \over 5}  {\beta \over k} b_2
 \left({\dot \eta(\beta,t)}
+ {{\dot h(\beta,t)} \over 6} \right) ~.
\eqno (2.6o)
$$
Accordingly, when open models show negative TE correlations at low $l$,
we expect something similar in flat RP models. On the contrary, 
models like SUGRA, with a cosmic acceleration closer to 
$\Lambda$CDM models, are not expected to give negative TE correlation.

Let us then discuss how the likelihood distribution can be obtained 
from model spectra, taking into account that the number 
of pixels for anisotropy and polarization ($N_T$ and $N_P$) in our 
(artificial) data are different. Let $T_j$ be the anisotropy data
measured in $N_T$ pixels and $Q_j$ and $U_j$ be the Stokes parameters
in $N_P$ pixels. In general, let be ${\bf x} \equiv 
(T_1,.....,T_{N_T}, Q_1,.....,Q_{N_P}, U_1,.....,U_{N_P})$.
${\bf x}$ is a vector of $N_s = N_T + 2 N_P$ 
components, defining an observed state of anisotropy and 
polarization. Once a model is assigned, the $C_l$ are 
uniquely determined. Passing to a data vector ${\bf x}$, instead,
amounts to performing a model realization. {\it Vice--versa},
once the $N_s$ component data vector ${\bf d}$ is given, the model
is not uniquely fixed.

   \begin{figure}  
   \vspace*{-5.cm}  
   \begin{center}
   \epsfig{figure=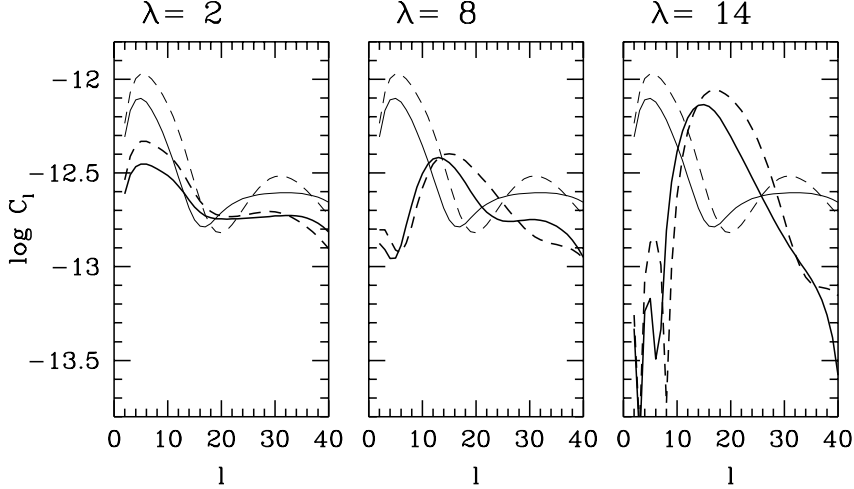,width=12.5cm}  
   \end{center}
   \vspace*{-4.cm}  
   \caption{ TE correlation spectra [$C_l = l(l+1) |C_l^{TE}|/4\pi$] 
for $\Lambda$CDM (the same  in the three plots) and RP.
Solid (dashed) lines refer to
$\tau = 0.14$ ($0.20$). In the third plot, low--l peaks are 
negative.
   } 
   \end{figure} 

Then, the likelihood of a model, whose angular spectra are $C_l$,
when the data ${\bf d}$ are observed, reads
$$
{L}({\bf d}|C^A_l) \propto [\, \det {\bf M}\, ]^{-{1 \over 2}} \exp 
\big[ -{1 \over 2}{\bf d^T} {\bf M}^{-1} {\bf d} \big] ~.
\eqno (2.7)
$$
The main ingredient of $L$ is the correlation matrix 
${\bf M_{ij}} = \langle {\bf x}^T_i {\bf x}_j \rangle = {\bf S}_{ij} 
+ {\bf N}_{ij}$; here ${\bf S}_{ij} $ is the signal term and
${\bf N}_{ij}$ is due to the noise. The components ${\bf M_{ij}}$
yield the expected correlation between the $i$th and $j$th elements of 
data vectors ${\bf x}$ corresponding to particular choices of
$C_{l}$ [10].  
The construction of the noise term 
is simpler, as we expect no noise correlation, and the matrix 
${\bf N}_{ij} = \delta_{ij}\,{\sigma_{T,pix}}^2$ (for $i = 1,...,N_T$) 
and ${\bf N}_{ij} = \delta_{ij}\,{\sigma_{P,pix}}^2$ (for $i = 
N_T+1,...,N_s$) is diagonal.

\section{Angular spectra and likelihood: results.}
In Fig.~1 we show the spectra $C_l^{TE}$ for $\Lambda$CDM and RP models,
if $\tau=0.14$ and $\tau=0.20$ (WMAP[11] suggests that $\tau = 0.17 \pm 0.04$).
The value of $\lambda = \log_{10}(\Lambda/{\rm GeV})$ is given in top of 
the frames. Low--$l$ anticorrelation is found only for $\lambda >\sim 10$ and
is shown in the third plot. The differences between $\Lambda$CDM and RP
increase for growing $\lambda$, but are already significant even for 
$\lambda=2$.

At low $l$, cosmic variance must be taken into account,
aside of instrumental variance. We then perform a large number (1000)
of realizations of RP sky models for $\lambda =$2, 5 and 8, for
$\tau=0.14$ and 0.20. In Fig.~2 (3) we report report results
for $\tau=0.14$ (0.20).
The values of $\lambda$ are shown in top of each plot.
Lower and upper plots correspond to $\sigma=2$ and 0.2$\, \mu$K,
respectively.  The former
value approaches WMAP and SPOrt expectations; the latter
value might be approached by the PLANCK [12] experiment.

   \begin{figure}  
   \vspace*{-2.cm}  
   \begin{center}
   \epsfig{figure=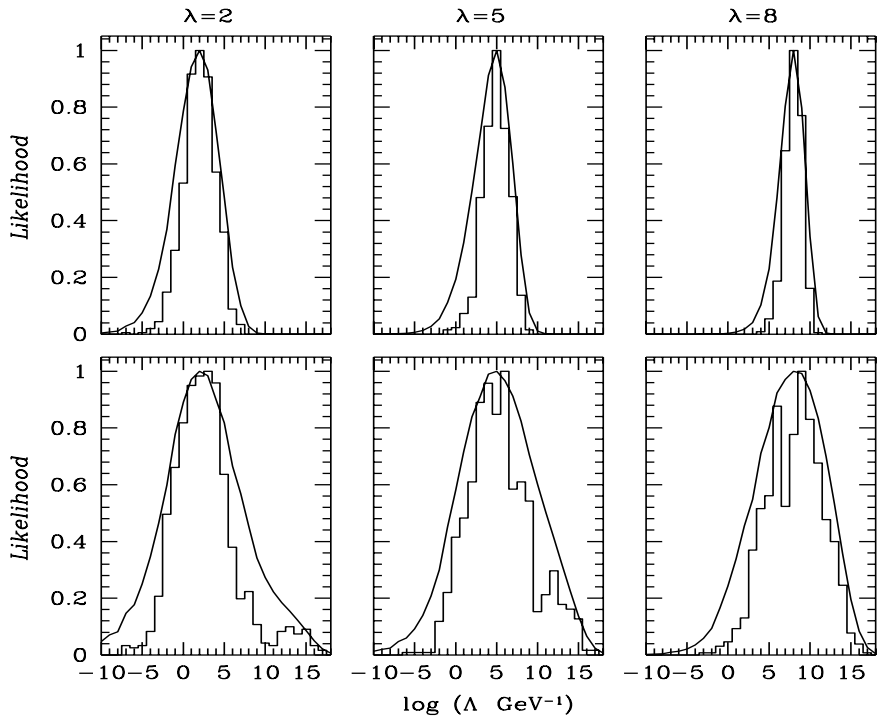,width=12.5cm,height=10.cm}  
   \end{center}
   \vspace*{-1.cm}  
   \caption{ Histograms give the distribution of peak likelihood. 
Continuous curves give the likelihood distribution averaged
over realizations (arbitrary but equal normalization).
Pixel noise is 0.2$\, \mu$K in the upper panels and 2$\, \mu$K in the
lower panels; all plots refer to $\tau = 0.14$.} 
   \end{figure} 

   \begin{figure}  
   \vspace*{-1.5cm}  
   \begin{center}
   \epsfig{figure=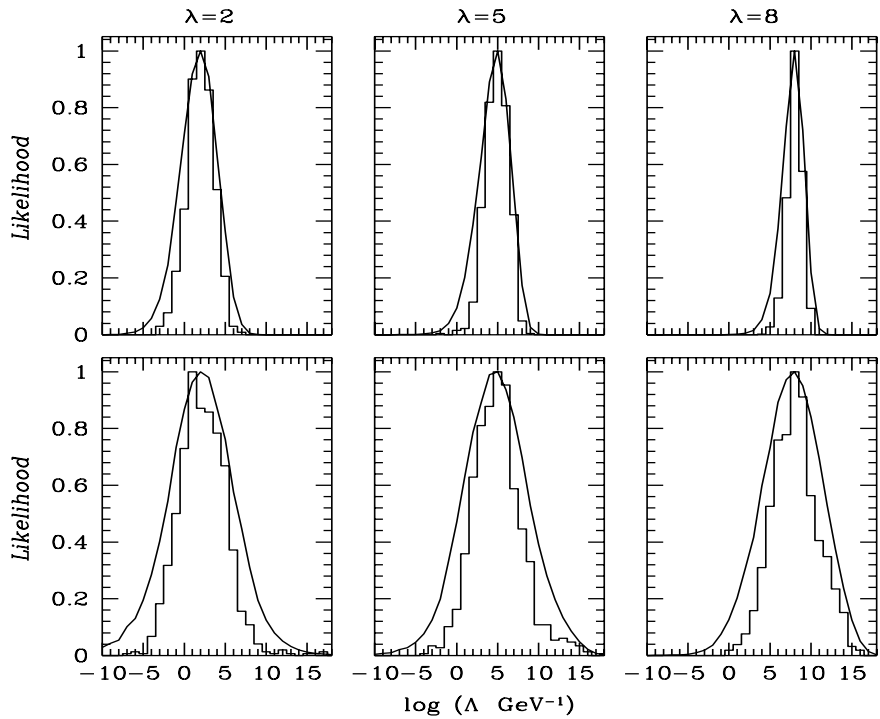,width=12.5cm,height=10.cm}  
   \end{center}
   \vspace*{-1.cm}  
   \caption{ The same as Fig.~2, but for an optical depth of $\tau =0.20$.} 
   \end{figure}

\section{Conclusions} 
These figures show that, as expected, RP models are more
easily distinguishable from $\Lambda$CDM for lower $\sigma$ 
and higher $\lambda$. We can assume that a RP model gives a signal
different from $\Lambda$CDM when $\lambda > 0$ is detected.
However, even in the less favorable case considered, when $\sigma 
= 2\, \mu$K, $\tau = 0.14$ and $\lambda = 2$, the peak likelihood is at
$\lambda > 0$ in $\sim 72\, \%$ of cases. For $\lambda = 5$, this
fraction reaches $\sim 91\, \%$. Likelihood distributions,
averaged both over cosmic and instrimental variances,
tell us that RP models begin to be distinguishable from
$\Lambda$CDM, at 1--$\sigma$ level, for $\lambda \sim 5$
and $\tau = 0.20$, if $\sigma=2\, \mu$K. For $\sigma=.2\, \mu$K,
instead, this is already true for $\lambda = 2$.
See the figures for further details.

A general conclusion that can be drawn from this analysis is that
cosmic variance is far from being a serious limit to model
parameter detection from large angle spectral analysis, at 
the present sensitivity levels. Even for a sensitivity
improved by a factor 10, hystograms are still contained
inside the curve, showing that there is a relevant
space for further improvements.

\vfill 
\end{document}